\documentclass[a4paper,useAMS,usenatbib]{mn2e}

\usepackage{times}
\usepackage{graphicx}
\usepackage{psfig}

\begin{document}

\title[The dark energy equation of state]{The dark energy equation of state}

\author[Usmani, Ghosh, Mukhopadhyay, Ray \& Ray]
{A. A. Usmani$^{1}$\thanks{E-mail: anisul@iucaa.ernet.in}, P. P.
Ghosh$^{2}$, Utpal Mukhopadhyay$^{3}$, P. C. Ray$^{4}$, Saibal
Ray$^{5}$\thanks{E-mail: saibal@iucaa.ernet.in}\\ $^{1}$Department
of Physics, Aligarh Muslim University, Aligarh 202 002, Uttar
Pradesh, India \\ $^2$Tara Brahmamoyee Vidyamandir, Matripalli,
Shyamnagar 743 127, North 24 Parganas, West Bengal, India\\
$^3$Satyabharati Vidyapith, Nabapalli, North 24 Parganas, Kolkata
700 126, West Bengal, India\\ $^4$Department of Mathematics,
Government College of Engineering and Leather Technology, Kolkata
700 098, West Bengal, India\\ $^{5}$Department of Physics, Barasat
Government College, North 24 Parganas, Kolkata 700 124, West
Bengal, India }

\date{Accepted . Received ; in original form }

\pagerange{\pageref{firstpage}--\pageref{lastpage}} \pubyear{2008}

\maketitle

\begin{abstract}
We perform a study of cosmic evolution with an equation of state
parameter $\omega(t)=\omega_0+\omega_1(t\dot H/H)$ by selecting a
phenomenological $\Lambda$ model of the form, $\dot\Lambda\sim
H^3$. This simple proposition explains both linearly expanding and
inflationary Universes with a single set of equations. We notice
that the inflation leads to a scaling in the equation of state
parameter, $\omega(t)$, and hence in equation of state. In this
approach, one of its two parameters have been pin pointed and the
other have been delineated. It has been possible to show a
connection between dark energy and Higgs-Boson.
\end{abstract}


\section{Introduction}
Cosmological research is mainly concerned with time (and in some
cases space as well) evolution of various physical parameters like
scale factor, Hubble parameter, matter-energy density etc. Along
with these parameters, in recent years a new physical entity
$\Lambda$ has resurrected in the foreground of cosmology. In fact,
$\Lambda$ has become an essential part of the field equations of
Einstein after some observational results
\citep{Riess1998,Perlmutter1999} indicated towards an accelerating
Universe. It is believed by most of the physicists that the
cosmological parameter $\Lambda$ is responsible for driving the
present acceleration because it can exert negative pressure.
Moreover, due to some fine-tuning problem (known as cosmological
constant problem), $\Lambda$ is regarded as a variable quantity
rather than a constant.

Now, in order to specify exact time-dependence of the unknown
physical quantities including $\Lambda$, one has to take recourse
of a relationship between cosmic pressure $p$ and matter-energy
density $\rho$ involving the equation of state parameter $\omega$.
Mathematically speaking, one variable quantity can depend on the
product of two other variable quantities. So,  one may construct
$\omega$ as a function of time, red-shift or scale factor
\citep{Chevron2000,Zhuravlev2001,Peebles2003}. In fact, values of
$\omega$ at different stages of cosmic evolution suggest that it
may evolve with time. As an instance, for the present
pressure-less Universe, the value of $\omega$ is considered as
zero, whereas its value was $1/3$ in the early radiation dominated
Universe. However, it is convenient to consider $\omega$ as a
constant quantity because observational data can hardly
distinguish between a varying and a constant equation of state
\citep{Kujat2002,Bartelmann2005}. Here some useful limits on
$\omega$ as appeared from SNIa data are $-1.67 <\omega < -0.62$
\citep{Knop2003} whereas refined values were indicated by the
combined SNIa data (with CMB anisotropy) and galaxy clustering
statistics which is $-1.33 < \omega < -0.79$ \citep{Tegmark2004}.

As stated above, $\omega$ may have a functional relationship with
scale factor or cosmological redshift. In connection to redshift
it may depend linearly, $\omega(z) = \omega_o + {\omega}^{\prime}
z$, where ${\omega}^{\prime}= (d\omega/dz)_{z=0}$
\citep{Huterer2001,Weller2002} or it may have a non-linear
relationship as $\omega(z) = \omega_o + {\omega}_1 z/(1+z)$
\citep{Polarski2001,Linder2003}.  This suggests for a simple form
\begin{eqnarray}
\label{omegaq} \omega(t)= \omega_0 + \omega_1 (t \dot H/H),
\end{eqnarray}
which  has got an explicit time dependence that disappears with the
condition, $t\dot H =H$.

Using above proposition, we explore the physical features of
different stages of cosmic evolution, viz., linearly expanding and
inflationary Universes. For this, a phenomenological $\Lambda$
model is selected to solve the Einstein field equations. There are
mathematically motivated
works~\citep{Ray2007,Mukhopadhyay2005,Mukhopadhyay2007a,Mukhopadhyay2007b},
wherein several phenomenological $\Lambda$ models have been
investigated for time-dependent $\omega$.

\section{Field Equations for a static spherically symmetric source}
The Einstein field equations are
\begin{eqnarray}
\label{eq1}
R^{ij}-\frac{1}{2}Rg^{ij}= -8\pi G\left[T^{ij}-\frac{\Lambda}{8\pi
G}g^{ij}\right],
\end{eqnarray}
where $\Lambda$ is the time-dependent cosmological term with
vacuum velocity of light being unity in relativistic units.

From equation~(\ref{eq1}) and Robertson-Walker metric, we get the
Friedmann and Raychaudhuri equations, respectively
\begin{eqnarray}
\label{eq3}
3H^2+\frac{3k}{a^2} &=& 8\pi G\rho+\Lambda,\\
\label{eq4}
3H^2+3\dot H &=& -4\pi G(\rho+3p)+\Lambda.
\end{eqnarray}
Here, $a=a(t)$ is the scale factor and  $k$ is the curvature
constant which assumes  values $-1$, $0$ and $+1$ for open, flat
and closed models of the Universe respectively. Also, $H=\dot a/a$
is the Hubble parameter and $G$, $\rho$, $p$ are the gravitational
constant, matter energy density and pressure respectively.
However, the generalized energy conservation law for variable $G$
and $\Lambda$ is derived by~\citet{Shapiro2005} using
Renormalization Group Theory and also by~\citet{Vereschagin2006}
using a formula of~\citet{Gurzadyan2003}.
 For variable $\Lambda$ and constant $G$,
the generalized conservation law reduces to the form
\begin{eqnarray}
\label{eq5}
\dot\rho+3(p+\rho)H= -\dot\Lambda/(8\pi G).
\end{eqnarray}

\section{Cosmological models for variable equation of state parameter}
The barotropic equation of state which relates the pressure and
density of the physical system is given by
\begin{eqnarray}
\label{eqstat}
p= \omega\rho.
\end{eqnarray}
Using this equation with equation~(\ref{eq5}), we arrive at
\begin{eqnarray}
\label{eq7}
8\pi G\dot\rho+\dot\Lambda= -24\pi G(1+\omega)\rho H.
\end{eqnarray}
For a flat Universe($k=0$), equation~(\ref{eq3}) yields
\begin{eqnarray}
\label{eq8}
-4\pi G\rho= \dot H/(1+\omega).
\end{eqnarray}

The equivalence of three phenomenological $\Lambda$-models (viz.,
$\Lambda \sim (\dot a/a)^2$, $\Lambda \sim \ddot a/a$ and $\Lambda
\sim \rho$) have been studied in detail by~\citet{Ray2007} for
constant $\omega$. So, it is reasonable to study a
variable-$\Lambda$ model with a variable $\omega$. Let us,
therefore, use the {\it ansatz} $\dot\Lambda \propto H^3$, so that
\begin{eqnarray}
\label{eq9}
\dot\Lambda= AH^3.
\end{eqnarray}
This {\it ansatz} may find realization in the framework of self
consistent inflation model~\citep{Dymnikova2000,Dymnikova2001}, in
which time-variation of $\Lambda$ is determined by the rate of
Bose condensate evaporation~\citep{Dymnikova2000} with $A \sim
(m_B/m_P)^2$ (where $m_B$ is the mass of bosons and $m_{P}$ is the
Planck mass).

From equations~(\ref{eq4}),(\ref{eqstat}),(\ref{eq8}) and
(\ref{eq9}), we get
\begin{eqnarray}
\label{eq10}
\frac{2}{(1+\omega)H^3}\frac{d^2H}{dt^2}+\frac{6}{H^2}\frac{dH}{dt}=A.
\end{eqnarray}
With $dH/dt=\dot H$, equation (\ref{eq10}) reduces to
\begin{eqnarray}
\label{meq} \frac{d\dot H}{dH}+3(1+\omega)H=
\frac{A(1+\omega)H^3}{2\dot H}.
\end{eqnarray}

We would now show, how does these field equations used in
conjunction with our proposition (equation~\ref{omegaq})
encorporate both linearly expanding and inflationary Universes.

\section{Linearly expanding Universe}

We consider a situation in which our Universe started expanding
linearly \citep{Crane1979,Azuma1982,Calzetta1983} since its very
beginning at a rate $\dot H=dH/dt$ with $H(t=0)=0$ at the point of
singularity. Thus at a later time $t>0$, the observable $H(t)$
would be determined by the relation, $H(t) = t\dot H$. The $\dot
H$ is the present value of $H$ divided by the age of the Universe.
In this case, equation~(\ref{meq}) reduces to
\begin{eqnarray}
\label{lineq} \frac{d\dot H}{dH} +3(1+W)H =\frac{A(1+W) H^3}{2\dot
H}
\end{eqnarray}
where $W=\omega_0+\omega_1$.

Solution set for the  differential equation~(\ref{lineq}) in
connection to different physical parameters is given below,
\begin{eqnarray}
\label{ac} a(t)&=& C (Et+D)^{1/E},\\
\label{hc} H(t)&=&\frac{1}{Et+D},\\
\label{wc} \omega(t)&=&\omega_0+\omega_1\left(\frac{1}{1+\frac{D}{Et}}\right),\\
\label{rhoc} \rho(t)&=&\frac{E}{4\pi G(Et+D)^2(1+\omega(t))},\\
\label{pc} p(t)&=&\omega(t)\rho(t),\\
\label{lamc} \Lambda(t)&=&- \frac{A}{2E(Et+D)^2}.
\end{eqnarray}
Here, $C$ and $D$ are integration  constants and $E$ reads as
\begin{equation}
\label{ee} E=\left[3(1+W)+\sqrt{9(1+W)^2+4A(1+W)}\right]/4.
\end{equation}
With the fact that $A<<W$, we may neglect the term involving $A$
in the above equation, which would yield $E\approx 3(1+W)/2$.
However, this would amount to be neglecting r.h.s term, ${A(1+W)
H^3}/{2\dot H}$, of equation~(\ref{lineq}), which suggests that
the effect of this term is small. It is also obvious from
equation~(\ref{lineq}) that this term matters only at an early
stage of the evolution of the Universe where $H \sim A $. However,
at this regime quantum effects become important and hence are of
no relevance in our general relativistic approach.

With the consideration, $H(t)=\dot H$, equation~(\ref{omegaq})
does not involve any explicit time dependence.  So is
equation~(\ref{wc}) provided $D=0$. We notice that with $E=1$ and
integration constants $D=0$ and $C=1$, equation~(\ref{ac}) becomes
a perfect example of a linearly expanding Robertson-Walker
Universe, $a(t)=t$. However, $E=1$ suggests a value
$W=w_0+w_1=\omega(t)=-1/3$, which is well above the minimum limit
of $\omega(t)$ i.e. $- 0.79$. We would see it later that inflation
scales it to a lower value. From equation~(\ref{hc}), deceleration
parameter, $q$, is deduced to be $q=E-1$, which thus is zero for
such a linearly expanding Universe.
\begin{figure}
\includegraphics{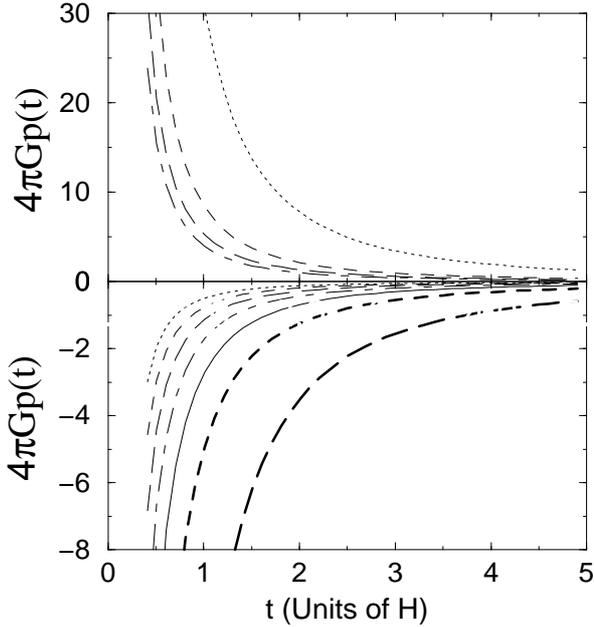}
\caption{\label{fig1} The upper panel represents $1+\omega(t)<0$.
In this panel, dotted, dashed, long-dashed, and chain curves
correspond to $\omega_1=-0.7, -0.8, -0.9$ and $1.0$, respectively.
In the lower panel representing $1+\omega(t)>0$, same curves
correspond to $\omega_1=0.0, -0.1, -0.2$ and $-0.3$, respectively.
The solid, thick dashed and thick long-dashed lines represent
$\omega_1=-0.4, -0.5$ and $-0.6$, respectively. For all these
$\omega_0$ is taken to be $-1/3$. }
\end{figure}
\begin{figure}
\includegraphics{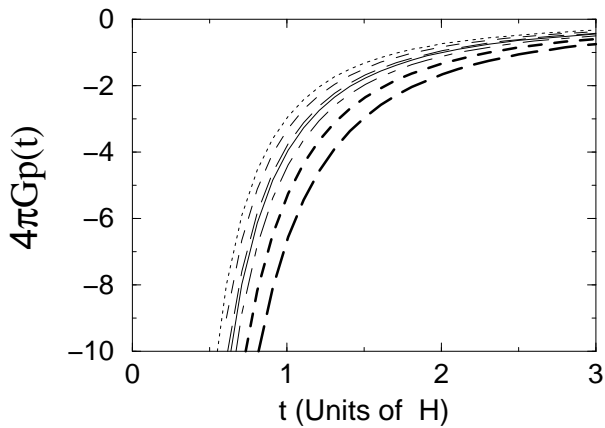}
\caption{\label{fig2} The dotted, dashed, long-dashed, chain and
solid curves represent $\omega_0=-0.1, -0.2, -0.3, -0.4$ and
$-1/3$, respectively. The thick dashed and thick long-dashed lines represent
$\omega=-0.5$ and -0.6, respectively.
For all these, $\omega_1$ is adjusted using
$\omega(t)=\omega_0+\omega_1=-0.8$}
\end{figure}
\begin{figure}
\includegraphics{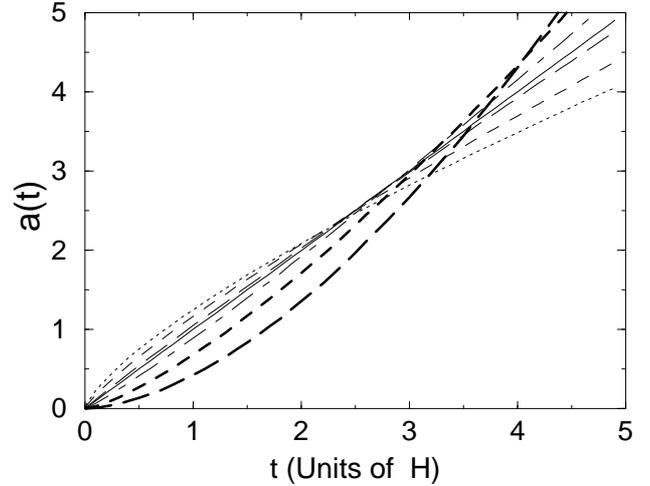}
\caption{\label{fig3} The scale factors for the curves shown in
Figure~\ref{fig2}.  }
\end{figure}

\section{Inflationary Universe}
We now consider a physical situation in which our Universe
initially inflated non-linearly up to a certain value of time
$t=t_0 << 1$ second \citep{Guth1981,Linde1982,Albrecht1982}. Since
this time onward the expansion of the Universe is assumed to be
quite linear, which is described by the rate $\dot H=dH/d\tau$.
Here $\tau$ is the measure of the time from $t=t_0$. This leads to
a translation in $H$ such that $H(t=t_0+\tau)=H(t_0)+\tau\dot H$.
We assume that inflation has led to a condition $H(t_0)>>\tau\dot
H$, which implies that $H(t)=H(t_0+\tau)>>\tau\dot H$. With the
consideration that the period of inflation has been very very
brief compared to the age of the Universe, we may write $t\approx
t_0+\tau$ and $\dot H= dH/dt\approx dH/d\tau$. However, the value
of $\dot H$ would be different from the previous case of linearly
expanding Universe. Under these conditions, equation~(\ref{meq})
reduces to
\begin{equation}
\label{infeq} \frac{d\dot
H}{dH}+3(1+\omega_0)H=\frac{A(1+\omega_0) H^3}{2\dot H}.
\end{equation}
If we substitute $W$ at the place of $\omega_0$ in
equation~(\ref{lineq}), we arrive at equation~(\ref{infeq}).  The
solution set obtained for the linearly expanding Universe is still
valid for the inflationary Universe provided we substitute
$\omega_0$ at the place of $W$ in equation~(\ref{ee}). This
scaling from $W$ to $\omega_0$ in equation~(\ref{ee}) may be
attributed to the adiabatic expansion of the Universe till time
$t_0$. The r.h.s. of equation~(\ref{infeq}) may be always
neglected in this case because $H$ is  evolved to a large value
compared to the values of $A$ during inflation.

With the consideration that $ A<< \omega_0$, we obtain
$\omega_0=-1/3$. Thus, the value
$\omega(t)=\omega_0+\omega_1=-1/3$ as obtained for linearly
expanding Universe now corresponds to
$\omega_0=\omega(t)-\omega_1=-1/3$ for an inflationary Universe.
Therefore, the values $\omega_0=-1/3$ and $\omega_1=0$ correspond
to previously discussed linearly expanding Universe  and a nonzero
value for $\omega_1$ represents inflationary Universe. Thus, we
notice a direct correlation between $\omega(t)$ and the inflation
of the Robertson-walker Universe, which is buried in the value of
the parameter $\omega_1$. With $\omega_0=-1/3$, the range of the
values $-1.0 <\omega_1<-0.46$ falls in the suggested range $-1.33
<\omega(t)<-0.79$.

We may invoke a time dependence in equation~(\ref{wc}) through
$D$. However, as mentioned earlier, data do not suggest any
significant explicit time dependence in $\omega(t)$, thus $D$ is
set to zero. The non-linearity in $a(t)$ may be invoked through
$\omega_0$ in $E$ by choosing a different  value for it other than
$-1/3$. Thus for a linear behaviour after inflation this value is
fixed to $-1/3$. The equation~(\ref{rhoc}) for $\rho$ is singular
at $1+\omega(t)=0$. So is equation~(\ref{pc}) for $p$, which has
been plotted in Figure~\ref{fig1}. For the negative pressure, as
required by the dark energy, it applies a constraint on
$\omega(t)$ such that $\omega(t)>-1$ or
$\omega_1=\omega(t)-\omega_0>-2/3$. We find a  range
$-2/3<\omega_1 <-0.46$ with $\omega_0=-1/3$.

\section{Discussion and remarks}

We have discussed two Universes: (i) a linearly expanding Universe
from its very beginning, (ii) and also the Universe like ours,
which has gone through an inflation at its very early stage
followed by a linear expansion later. We notice that these two
kind of Universes, which  are direct consequence of our
proposition (equation~\ref{omegaq}), are represented by the same
set of equations with a translational shift in the  equation of
state parameter in the latter case compared to the former. In both
the cases, $a(t)=1$ demands $E=1$, which applies a constraint on
the equation of state parameter.  For the inflationary Universe,
we have pin pointed $\omega_0=-1/3$ and have delineated the other
parameter with a range $-2/3 <\omega_1< -0.46$. We observe that
former is a special case of the latter with $\omega_0=-1/3$ and
$\omega_1$=0. Any other value of $\omega_0$ would invoke a
non-linear behaviour in $a(t)$ through $E$. The  effect of the
variation of $\omega_0$ on $p$ is presented in Figure~\ref{fig2}
for a constant $\omega=\omega_0+\omega_1=-0.80$ obtained by
adjusting $\omega_1$ accordingly. The $\omega_1$ has nothing to do
with $E$ and hence has nothing to do with $a(t)$. However, its
value is a measure of translation in $\omega$ due to inflation.
The equations for $\rho$ and $p$ involve $\omega$ and hence would
remain unchanged with its constant value. Thus, variations in
curves of Figure~\ref{fig2} is purely due to the variation in
$\omega_0$. The corresponding variations in $a(t)$ are shown in
Figure~\ref{fig3}.

A negligible value of $A$ is shown to be physically possible from
the viewpoint of cosmology and particle physics, which means the
absence of $\Lambda$ in the field equations. So, both from
physical and mathematical point of view the nullity of $\Lambda$
is achieved for the same $\Lambda$ model. Again, the expression of
$q$ in this case has a striking similarity with that
of~\citet{Ray2007}. This work suggests that in the late phase of
the Universe, where $t\dot H=H$, the equation of state parameter
behaves as a constant. Perhaps for this reason current data cannot
distinguish clearly between a time-dependent $\omega$ and a
constant one as pointed out by some workers
\citep{Kujat2002,Bartelmann2005}.

Separating the entire cosmic history into two phases, it has been
possible to derive the time-dependent expressions for the scale
factor and the other physical parameters of each phase. It has
been found that for inflationary phase, the deceleration parameter
$q$ depends on time whereas for the linearly expanding phase it is
constant, rather zero. This supports the opinion that $q$ has
changed during the course of
time~\citep{Riess2001,Amendola2003,Padmanabhan2003}.

\section*{Acknowledgments}
Authors (AAU and SR) are thankful to the authority of
Inter-University Centre for Astronomy and Astrophysics, Pune,
India for providing Associateship programme under which a part of
this work was carried out.

{}
\end{document}